\PassOptionsToPackage{utf8}{inputenc}
\documentclass[nocrop]{bioinfo}
\copyrightyear{2020} \pubyear{2020}
\usepackage{bm} 
\access{Advance Access Publication Date: 14 December 2020}
\appnotes{Manuscript Category}

\begin{document}
\firstpage{1}

\subtitle{Bioinformatics}

\title[SimpleChrome]{SimpleChrome: Encoding of Combinatorial Effects for Predicting Gene Expression}
\author[SimpleChrome]{Wei Cheng$^{\text{\sfb 1}}$, Ghulam Murtaza$^{\text{\sfb 1}}$, and Aaron Wang$^{\text{\sfb 1,}*}$}
\address{$^{\text{\sf 1}}$Computer Science Department, Brown University, Providence, RI, 02912, USA}

\corresp{$^\ast$To whom correspondence should be addressed.}

\history{}
\editor{}
\abstract{Due to recent breakthroughs in state-of-the-art DNA sequencing technology, genomics data sets have become ubiquitous. The emergence of large-scale data sets provides great opportunities for better understanding of genomics, especially gene regulation. Although each cell in the human body contains the same set of DNA information, gene expression controls the functions of these cells by either turning genes on or off, known as gene expression levels. There are two important factors that control the expression level of each gene: (1) Gene regulation such as histone modifications can directly regulate gene expression. (2) Neighboring genes that are functionally related to or interact with each other that can also affect gene expression level. Previous efforts have tried to address the former using Attention-based model. However, addressing the second problem requires the incorporation of all potentially related gene information into the model. Though modern machine learning and deep learning models have been able to capture gene expression signals when applied to moderately sized data, they have struggled to recover the underlying signals of the data due to the nature of the data's higher dimensionality. To remedy this issue, we present SimpleChrome, a deep learning model that learns the latent histone modification representations of genes. The features learned from the model allow us to better understand the combinatorial effects of cross-gene interactions and direct gene regulation on the target gene expression. The results of this paper show outstanding improvements on the predictive capabilities of downstream models and greatly relaxes the need for a large data set to learn a robust, generalized neural network. These results have immediate downstream effects in epigenomics research and drug development.\\
\textbf{Supplementary Information:} Code and data is available at \href{https://github.com/aaronwangj/SimpleChrome}{https://github.com/aaronwangj/SimpleChrome}\\
\textbf{Contact:} \href{aaronjwang@brown.edu}{aaronjwang@brown.edu}
}

\maketitle

\section{Introduction}
The human body has thousands of different cell types, ranging from muscle cells to brain cells, and even though these cells contain the same DNA, their roles differ drastically due to the differential expression levels of different genes within cells, leading to downstream changes in the cell's overall functions. These changes may affect which genes in the cell are expressed.

One critical factor of gene expression is the modification of histone proteins, bead-like structures that provide support to chromosomes. The cell uses these histone proteins to organize the DNA in a condensed state, controlling which parts of the DNA are "exposed" and thus allowed to express. Histone proteins are prone to chemical modification which may change which parts of the DNA are exposed and expressed. This is known as the Histone Code Hypothesis. Unlike genetic mutations, another source of gene expression modification, histone modifications have shown to be potentially reversible, a powerful juxtaposition that has advanced the development of a new generation of targeted drugs that cure diseases resulting from reversible aberrant histone modifications. 
Recent advancements in sequencing technologies have allowed researchers to quantify gene expression and histone modification, the latest and most comprehensive data set being REMC (Roadmap Epigenome Project). Initial studies to understand the combinatorial effects of histone modifications on gene expression experimentally validated that there exists a correlation between histone modifications and gene expression. Computational methods, most notably DeepChrome by \cite{deepchrome} and AttentiveChrome by \cite{AttentiveChrome}, that employ deep learning have shown to exceedingly outperform all previous machine learning based methods in learning complex combinatorial gene interactions. Despite the deep learning models showing superior predictive performance, these models only consider a limited amount of data around the Transcription Start Site (TSS) to perform their predictions, in contrast to the models' complex architectures with many parameters. Previous literature has shown that neighboring genes may also play important roles in determining the gene expression level, as they may interact with each other in a functionally related pathway. Unfortunately, including more potential candidate genes such as neighboring genes into a conventional deep learning model would severely increase the number of input features, hindering the power and computational sustainability of the model. To overcome this issue, we develop SimpleChrome and DeepNeighbors.
DeepNeighbors is trained in two steps. First, it utilizes unsupervised learning to derive a lower dimensional representation of histone modifications for each gene as new input. In the second step, the representation of the target gene or the fused representations of both target gene and its neighbor genes is fed into a simple model for predicting gene expression. SimpleChrome refers to the first step of the training only and does not include the neighboring genes for predicting gene expression. We show that the representations learned by SimpleChrome can successfully preserve useful information of the original data for predicting gene expression; we demonstrate with SimpleChrome that this can significantly reduce required sample size and model complexity for achieving competitive prediction performance. We also show that spatially neighboring genes do not contribute significantly for predicting gene expression, seen in the performance of DeepNeighbors.

\section{Related Work}
\subsection{Previous Methods}
Over the past decade, many researchers have developed machine learning and deep learning-based models to predict the effect of histone modifications (HMs) on gene expressions. These studies primarily use the REMC data set, which contains the intensity and location of HMs across the entire genome across 56 different cell types. Recent studies have attempted to model this either as a classification or a regression task. The studies that have attempted to model it as a classification task have applied a slew of different techniques, including linear regression by \cite{linear_regression}, SVMs by \cite{svm}, Random Forests by \cite{random_forest}, Rule-Based Learning by \cite{rule-based}, and deep learning by \cite{AttentiveChrome, deepchrome}. In the latter-most category, DeepChrome and AttentiveChrome are cell-specific gene expression prediction frameworks that outperform all previously published machine learning-based techniques. Whereas DeepChrome utilizes CNNs to capture the histone marks' combinatorial interactions and autonomously learn the latent biological interactions, AttentiveChrome uses a hierarchical attention-based model to understand the latent biological interactions between the HMs to predict gene expression. Similarly, other studies have tried to model the same problem as a regression task,including the SVR (Support Vector Regression) model by \cite{svm} and DeepDIFF (Attention Based). These models attempt to estimate the differential gene expressions across different cell lines based on the HM's difference around the TSS (Transcription Start Site) and TTS (Transcription Termination Site). 
Although these models achieve state-of-the-art performance, they are much more complex and do not account for any impacts of neighboring genes.
%Unfortunately, all of the previous methods have failed to account for the impact of Histone modifications at other genes. 

\subsection{Proposed ideas}
Recent works have used the combination of gated neural networks with dual attention networks to predict chromatin accessibility, as seen in \cite{attentive_gated_networks}. As suggested by the authors, the gated network structure may not only stabilize variances but also avoid vanishing and exploding gradients, leading to converging towards a better model. Multiple embedding modules were also proposed by \cite{DeepANF} to learn DNA representations, which were shown to have better performances compared to vanilla RNNs, or gated networks. Thus, learning more efficient embeddings could be a potential improvement over existing models. In recent years, there have been many studies proposed to use generative models to learn input embeddings since generative modeling captures the underlying mechanisms of how the data is generated, and could thus learn more useful information from them.  Some well-known models include Variational Autoencoders (VAEs), Generative Adversarial Networks (GANs), or their hybrid, Adversarial Autoencoders (AAEs). Previous studies show that using these models to first learn latent representations without labels and then use latent representations for downstream supervised tasks can significantly improve prediction accuracy, as seen by \cite{zeroshot, adversarial, semisupervised}. Therefore, we propose to combine generative models with previous techniques to improve performance.

\section{Methods}
\begin{figure*}[h]
  \centering
  \vspace{5pt}
  \includegraphics[scale=0.4]{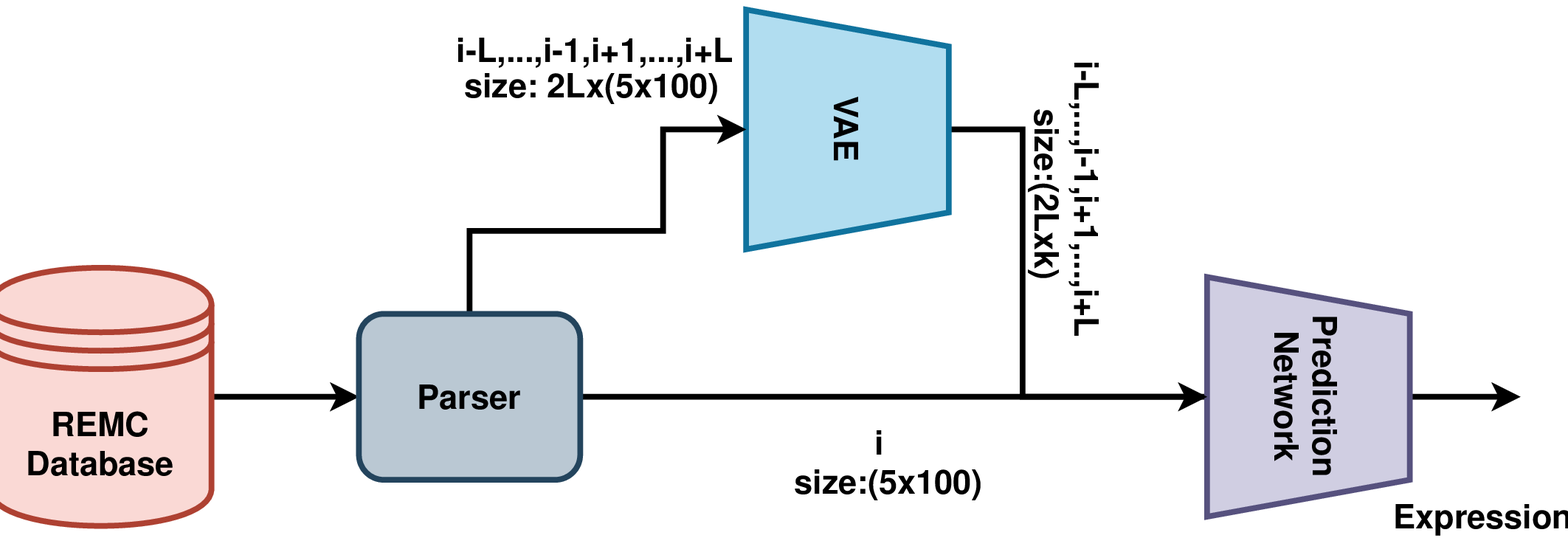}
  \caption{Training process of DeepNeighbor: 1. Data are preprocessed and the neighboring gene input matrices are transformed into lower dimensions using VAE and concatenated into a matrix $z$. 2. $z$ combines with $x$ from the target gene and is fed into an MLP model for predicting gene expression.}
  \label{fig:Schematic figure}
\end{figure*}

\begin{figure*}[h]
  \centering
%   \vspace{20pt}
  \includegraphics[scale=1]{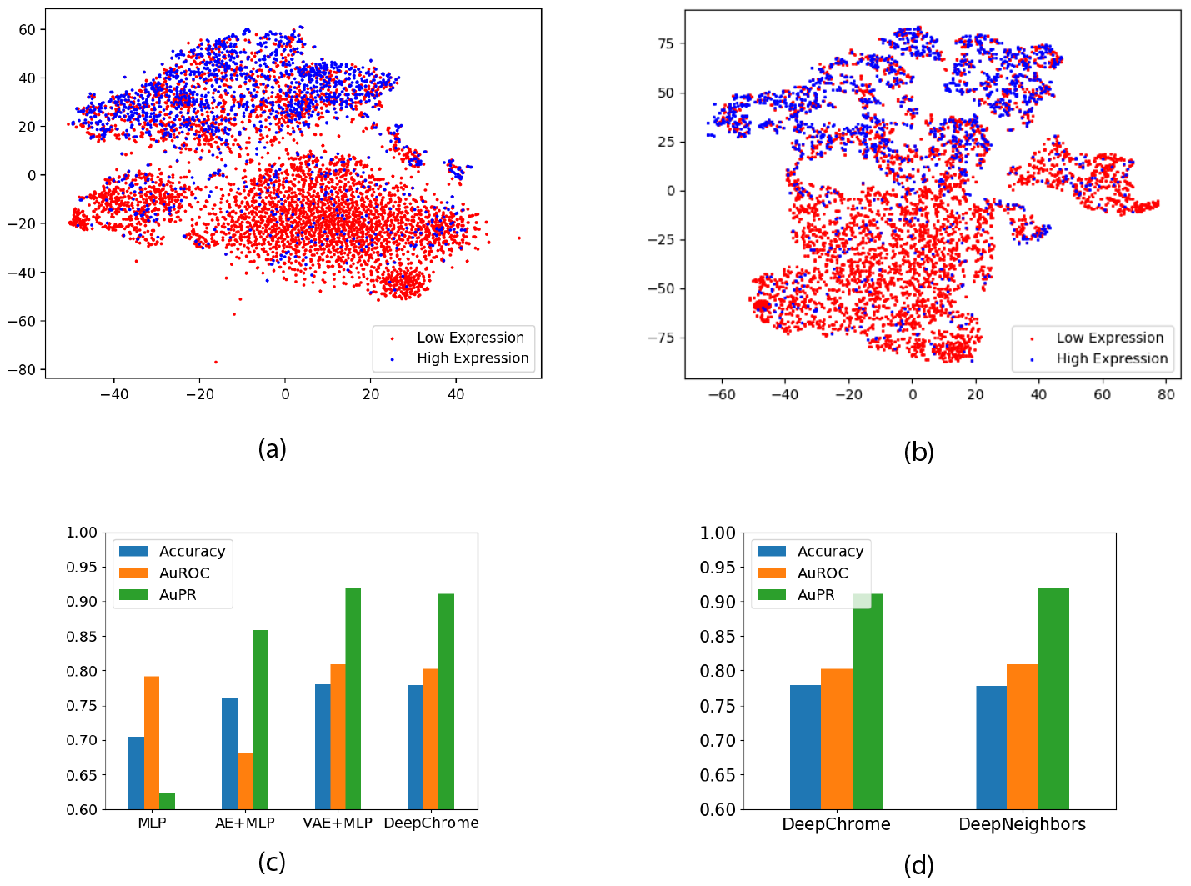}
  \caption{Figures (a) and (b) show the t-SNE plots of the encoder outputs. Figure (c) compares the performance of the MLP with different data inputs against DeepChrome. Figure (d) compares the predictive performance of SimpleChrome against DeepChrome}
  \label{f:tSNE}
\end{figure*}
\subsection{Representation Learning and Autoencoders}
We are given an input matrix $\bm{X}$ with dimension of $N \times P$, where $N$ denotes the number of samples and $P$ indicates the number of features. As $P$ becomes larger, input matrix $\bm{X}$ becomes harder to work with for two main reasons. Firstly, storing and parsing the data in memory as well as passing and processing the data with the models can be computational expensive. Secondly, the objective of the model becomes harder to optimize and requires significantly larger $N$ to derive an optimal solution. Hence, we are interested in converting $\bm{X}$ into $\bm{Z}$ with dimension of $N \times K$ where $K<<P$. Each column of $\bm{Z}$ is a lower dimension representation of the original data $\bm{X}$. Since the representations are supposed to preserve the essence of the original input and thus discard any redundant information within lower dimensions, downstream tasks such as predictions will greatly benefit.There exist multiple dimension reduction techniques in literature, specifically by \cite{scholz2008nonlinear, kasun2016dimension}, such as Principle Component Analysis (PCA), Tensor Factorization, Singular Value Decomposition (SVD) which are based on linear transformations; while linear models are easier to interpret and compute, they will lose non-linearity in the original data. Hence, in this paper, we consider non-linear models such as Autoencoders. 

An Autoencoder contains an encoder $f_{\theta}(\bm{x})$ and a decoder $g_{\phi}(\bm{z})$ parameterized with $\theta$ and $\phi$, respectively. The encoder $f_{\theta}(\bm{x})$ takes one sample of original data $\bm{x}$ as input and outputs a vector contains latent variables $\bm{z}$ in a lower dimension. The decoder $g_{\phi}(\bm{z})$ takes the latent variables $\bm{z}$ as input and outputs reconstructed data $\bm{x}'$ with the same dimensions of $\bm{x}$. The intuition behind the Autoencoder is that fitting latent variables $\bm{z}$ will enable better construction of $\bm{x}$. Thus, Autoencoders are often trained with the following objective,
\begin{equation}
    \sum_{i=1}^N||\bm{x}_i - g_{\phi}(f_{\theta}(\bm{x}_i))||^2 + \lambda \sum_{\bm{w} \in \{\bm{\theta}, \bm{\phi}\}} ||\bm{w}||_p,
\end{equation}
where $||\bullet||^2$ denotes the mean square error and $||\bm{w}||_p$ denotes regularization on the weights $\bm{w}$ of the model (with $p$ denotes $p$-th norm). As the encoder and the decoder contains nonlinear activation function such as ReLU, sigmoid, etc, Autoencoders are powerful tools for learning lower dimensional representation of raw, non-linear data $\bm{x}$.
\subsection{Generative Modeling and Variational Autoencoder}
Generative Modeling has drawn a lot of attention in the deep learning field for the past several years. Unlike most black-box methods in deep learning, generative models aim to learn the underlying distribution of data $P(\bm{x})$ which enables better understanding of how the data is generated, thus benefiting downstream tasks such as representation learning, generating new samples, etc. Two of the most popular methods are Generative Adversarial Networks (GANs) by \cite{goodfellow2014generative} and Variational Autoencoders (VAEs) by \cite{kingma2013auto}. While GANs were shown to generate high quality samples, the lack of theoretical support and the difficulty in adversarial learning hinders its utilities in the field, as seen by \cite{tolstikhin2017wasserstein}. In this paper, we focus on Variational Autoencoders (VAEs) which are easier to train and enjoy strong theoretical support from Bayesian inference.

Similar to the Autoencoder, a Variational Autoencoder has an encoder $f_{\theta}(\bm{x})$ and a decoder $g_{\phi}(\bm{z})$. An intuitive way to understand VAEs is such that while they aim to reconstruct inputs similar to vanilla Autoencoders, they also attempt to regularize the latent space $\bm{z}$ simultaneously. In brief, while the Autoencoders are assumed to only encode isolated points of $\bm{x}$ in the latent space, VAEs aim to derive a smooth latent space; more specifically, the goal of VAEs is to find a model that maximizes the marginal log-likelihood $\textnormal{log}p_{\theta}(\bm{x})$. While the marginal likelihood can be intractable in practice, Bayesian inference theory provides an alternative for practical optimization. According to Jensen's inequality, one can derive an Evidence Lower Bound (ELBO) of the marginal likelihood,
\begin{equation}
\begin{aligned}
&{\textnormal{log}}\,p_{\theta}(\mathbf{x}) \geq \mathbb{E} [{\textnormal{log}}\,p_{\theta}(\mathbf{x}\, | \,\mathbf{z})]-\,\text{KL}(q_{\phi}(\mathbf{z}\, | \,\mathbf{x})\,\|\, p_{\theta}(\mathbf{z})), \label{elbo}
\end{aligned}
\end{equation}
where $p_{\theta}(\mathbf{x}\, | \,\mathbf{z})$ denotes the conditional likelihood (given latent variables) of reconstructed data output by the decoder. One can interpret this by treating the output of the decoder as a probabilistic distribution (i.e. a Gaussian distribution for continuous sample); $q_{\phi}(\mathbf{z}\, | \,\mathbf{x})$ is the conditional posterior of latent variable $\bm{z}$ given an input $\bm{x}$ outputted by the encoder; $p_{\theta}(\mathbf{z})$ denotes the prior distribution which is commonly assumed to be standard Gaussian $\mathcal{N}(\bm{0}, \bm{I})$; the last term on the right hand side is the Kullback–Leibler divergence (relative entropy) between two distributions. When the prior and conditional posterior are both Gaussian, the KL term has a closed-form solution that can be directly trained with a gradient-based method. While the expectation of the likelihood term can be hard to derive in practice, \cite{kingma2013auto} introduce a Stochastic Gradient Variational Bayes (SGVB) method to approximate the expectation to overcome such an issue. Hence, modern VAEs can be easily trained with gradient descent methods. As the latent variables are appropriately `regularized' such that it maximizes the data probability, the latent variables should encode the inputs very well once the model converges.
\subsection{Two-Step Training for Predicting Gene Expression}
The training of the model can be dissected into two parts as shown in \ref{fig:Schematic figure}. In step one, we prepare relevant data and compress high dimensional data that may contain redundant information to lower dimensional representations. In step two, we consider two prediction models: i). A prediction model (i.e. Multilayer perceptron) that takes compressed data of the target gene as input. ii). A concatenation of target gene data with compressed neighbor genes data as input into the prediction model.
\subsubsection{Data Preparation and Dimension Reduction}
Following the work from \cite{deepchrome, AttentiveChrome}, we use five core Histone Modification marks for 56 different cell types derived from REMC datasbase; these five HM marks are H3K27me3, H3K36ME3, H3K4me1, H3K4me3 and H3K9ME3. For each cell type, of which there are 19,802 genes in total, we follow previous works and divide them into training (6601 genes), validation (6601 genes), and testing (6600 genes). In this study, we also consider using even smaller training sizes (100 genes or 1000 genes randomly sampled from 6601). For each gene, the 10000 bp (+/- 5000) region around the TSS (Transcription Start Site) is divided into bins of size 100. For each bin, we calculate the frequency of each Histone Modification, which gives a 5 by 100 matrix for the target gene $i$. Similar to AttentiveChrome and DeepChrome, we formulate the gene expression prediction as a binary classification task. Each target gene is represented as a 5 by 100 matrix $x_i$, the raw input for predicting the gene expression level. As a first step, we used either a Autoencoder or Variational Autoencoder architecture to learn a lower dimension of $k = \{2, 5, 10\}$ latent representation (we find that  $k=10$ optimizes performance) $z_i$. Then, we consider using either i)$z_i$, or ii). $x_i$ concatenated with $l=20$ neighbor $z_j$ as inputs into the prediction model. As a comparison, we also include prediction models that use $x_i$ as inputs such as DeepChrome. Due to the limitation of computational source, we only randomly picked three cell lines out of 56 and report them as the results.
% \subsubsection{Attention-based Join Prediction Model}
% With the previous setup, we have two input matrices $\bm{x}_i$ and $\bm{z}_i$. While each row of $\bm{x}_i$ stores the information of a specific histone modification, each row of the $\bm{z}_i$ contains the learned information of each gene. Similar to \cite{AttentiveChrome}, each row of $\bm{x}_i$ is fed into a separate LSTM network to learn spatial dependencies among its bin signals. Then an attention layer is applied to the output of each LSTM network and learned an embedding that extracted bin level information. The embedding is supposed to have the same dimension of each row of $\bm{z}_i$. Then $\bm{z}_i$ can be concatenated with the embedding as we assume in the dimension reduction phase, either bin or different HM information were already extracted and condensed into each row of $\bm{z}_i$. Next, the new matrix is fed into an attention layer for understanding the importance of each HM on the target gene as well as the neighbor genes. Finally, a fully connected layer with ReLU activation is used for output prediction.
\subsubsection{Baseline Models and Evaluation Metric}
We evaluate our model primarily based on predictive performance against well-known, state-of-the-art deep learning models. Since DeepChrome and AttentiveChrome outperform previous machine learning methods by a statistically significant margin, we focus our efforts on comparing our performance against DeepChrome and AttentiveChrome. However, while the focus of this work is to improve predictive capabilities, we choose DeepChrome as our sole baseline because as shows similar (or better) performance compared to AttentiveChrome, while still being interpretable.

The first baseline is DeepChrome, and it takes $x_i$ (5 by 100) as input matrix. We also included a two-layer MLP that takes $x_i$ as input as a comparison as well.

For VAEs and AEs, we used a two layer convolution encoder (64 and 128 filters) with batch normalization and ReLU activation, followed by two dense layers (one for learning means and one for learning variances). The decoder first uses a dense layer and then two convolution transpose layers with a final $tanh$ activation layer. The $tanh$ function compresses all data into the range of -1 and 1, which is then normalized. The AE is trained with MSE as the reconstruction loss and the VAE has an additional KL divergence for regularization loss. 

There are two main prediction models to evaluate, which are i). MLP, which uses $z_i$ as input. For all MLPs, we simply used a two layer architecture with 50 and 20 hidden neurons and ReLU activations. ii) a CNN similar to DeepChrome that will take $x_i$ and $z_j$ of 20 neighbored genes, in which neighbored genes are chosen by prioritizing those with the smallest Euclidean distances to the target gene measured using latent variables $z_i$. The CNN has the same convolution layer as DeepChrome that takes $x_i$ as input (50 filters with kernel size 10) and another convolution layer that takes all $z_j$ ($l = 20$ by $k = 10$) as input. Then, the output of these two convolution layers are concatenated and fed into a two layered MLP (same to DeepChrome) for prediction. All models are trained using the Adam optimizer with batch sizes of 100 and 10 epochs using training data or subsets of the training data.

\section{Results}
\begin{figure}[h]
  \centering
%   \vspace{20pt}
  \includegraphics[scale=0.25]{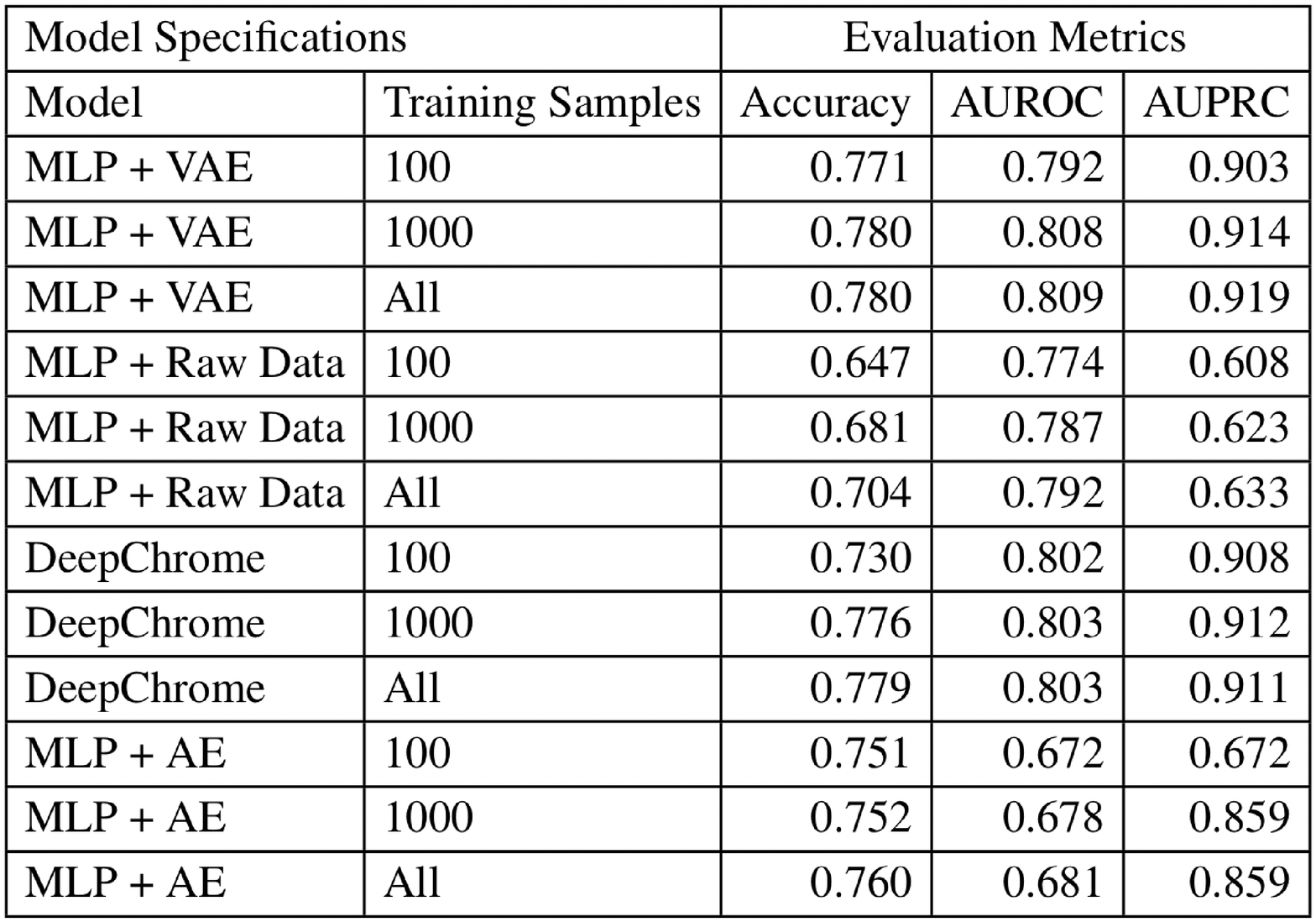}
  \caption{This table summarizes Accuracy, AuROC and AuPR scores of different models with varying sample sizes.}
  \label{f:sample_size}
\end{figure}
We systematically evaluate the following questions:
 a.) Whether VAEs or Autoencoders can encode useful information from raw data for predicting gene expression.
 b.) What the difference in performance is with DeepChrome and SimpleChrome.
c.) If models with encoded inputs learn faster and require less memory and data while maintaining equal performance.
% \begin{itemize}
%     \item a). Whether VAE or Autoencoder can encode useful information from raw data for predicting gene expression.
%      \item b). Performance of DeepChrome vs SimpleChrome (Its a fair starting point because attentive chrome improves on the interpretabiliy rather than performance).   
%     \item c). Models with encoded input learn faster and require less memory and data while maintaining equal performance. (Better for learning on huge data sets) 
% \end{itemize}

To qualitatively evaluate (a), we use tSNE to visualize the clustering of all the genes in the test set. We first run tSNE using flattened $x_i$; the plot is shown in Fig.\ref{f:tSNE}(a), which shows that genes with high expression and low expression seem to form two distinct clusters. We next run tSNE on $z_i$. We hypothesize that if VAEs or AEs can preserve useful information for predicting gene expression, then the tSNE results should preserve the distinct clusters. Indeed, as shown in Fig.\ref{f:tSNE}(b), one can see two clusters that are at least or even more distinct compared to Fig.\ref{f:tSNE}(a).
As a quantitative evaluation, we compared the performance of prediction (measured by accuracy, AUROC, AUPR) of DeepChrome using $x_i$ as input, MLP using $x_i$ as input, with MLP using $z_i$ from AE as input and MLP using $z_i$ from VAE as input. As shown in Fig.\ref{f:tSNE}(c). DeepChrome and MLP utilizing latent variables $z_i$ from VAE have similar performances overall while the AUPR of the latter is slightly better. Hence, we conclude that VAEs can better encode useful information than AEs.

To evaluate (b), we evaluate whether concatenating neighboring genetic information can better improve performance. Unfortunately, we did not see any improvement in performance overall, as the performance is statistically insignificant when compared to DeepChrome.

We found the most exciting results when evaluating (c). As introduced before, we hypothesized that generative modeling may have better performance when the sample size is limited. Thus, we sub-sampled 100 and 1000 samples for training all the previous models. The results are summarized in Fig.\ref{f:sample_size}. Surprisingly, we found out that MLP + VAE has similar and even better power when trained on only 100 or 1000 samples compared with DeepChrome using all 6601 training samples. The MLP model, with only 2 layers of 50 and 20 neurons, is substantially simpler and thus trains significantly faster compared to DeepChrome , yet maintained rivalling performance with less training data. It should be noted that we tested and recorded the training time speed of the full data set on the models using a MacBook Pro. The MLP model with latent variables takes less than 10 seconds, while DeepChrome takes one minute.
\section{Conclusion}
In this paper, we present a joint model system which attempts to include the histone modifications of spatially neighbored genes to improve the predictive capability in gene expression of existing state-of-the art deep learning models. However, in our preliminary evaluation, we find that including the spatially neighboring genetic information brings at most negligible improvement in the predictive power of the deep learning models. However, we show that our method of utilizing VAEs to encode the data set significantly improves the predictive capabilities of downstream models and greatly relaxes any requirements of a large data set to learn a robust, generalized model.

As an extension of our work, we intend to look at functionally relevant genes rather than solely spatially adjacent genes. This novel representation of the cell would allow our model to potentially improve its predictive capabilities when paired with an attention-style mechanism to provide insights into how and which proteins regulate gene expression. Furthermore, our training and evaluations were completed on a single cell line (E004) due to the limited timeline of the project, and thus future work could experiment on other cell lines.
\section*{Acknowledgements}
This work was conducted under Ritambhara Singh as part of the final course project for the graduate-level course Deep Learning in Genomics taught by Singh at Brown University. We appreciate the feedback from the instructor and our fellow students during the project's development.

\bibliographystyle{natbib}
\bibliography{document.bib}

% \begin{thebibliography}{}

% \bibitem[Bofelli {\it et~al}., 2000]{Boffelli03}
% Bofelli,F., Name2, Name3 (2003) Article title, {\it Journal Name}, {\bf 199}, 133-154.

% \bibitem[Bag {\it et~al}., 2001]{Bag01}
% Bag,M., Name2, Name3 (2001) Article title, {\it Journal Name}, {\bf 99}, 33-54.

% \bibitem[Yoo \textit{et~al}., 2003]{Yoo03}
% Yoo,M.S. \textit{et~al}. (2003) Oxidative stress regulated genes
% in nigral dopaminergic neurnol cell: correlation with the known
% pathology in Parkinson's disease. \textit{Brain Res. Mol. Brain
% Res.}, \textbf{110}(Suppl. 1), 76--84.

% \bibitem[Lehmann, 1986]{Leh86}
% Lehmann,E.L. (1986) Chapter title. \textit{Book Title}. Vol.~1, 2nd edn. Springer-Verlag, New York.

% \bibitem[Crenshaw and Jones, 2003]{Cre03}
% Crenshaw, B.,III, and Jones, W.B.,Jr (2003) The future of clinical
% cancer management: one tumor, one chip. \textit{Bioinformatics},
% doi:10.1093/bioinformatics/btn000.

% \bibitem[Auhtor \textit{et~al}. (2000)]{Aut00}
% Auhtor,A.B. \textit{et~al}. (2000) Chapter title. In Smith, A.C.
% (ed.), \textit{Book Title}, 2nd edn. Publisher, Location, Vol. 1, pp.
% ???--???.

% \bibitem[Bardet, 1920]{Bar20}
% Bardet, G. (1920) Sur un syndrome d'obesite infantile avec
% polydactylie et retinite pigmentaire (contribution a l'etude des
% formes cliniques de l'obesite hypophysaire). PhD Thesis, name of
% institution, Paris, France.

% \end{thebibliography}
\end{document}